\documentclass[twocolumn,showpacs,preprintnumbers,amsmath,amssymb]{revtex4}
\usepackage{graphicx} \usepackage{subfigure} \usepackage{amsbsy}
\begin{document}

\bibliographystyle{apsrev}

\title{Two-photon detuning and decoherence in cavity electromagnetically
  induced transparency for quantized fields}  \author{P. Barberis-Blostein}
\affiliation{Instituto nacional de astrof\'{i}sica \'optica y
  electr\'onica, Luis Enrique Erro 1, Tonantzintla, Puebla, M\'exico. }
\date{today}
\begin{abstract}
  The interaction of a quantized field with three-level atoms in $\Lambda$
  configuration inside a two-mode cavity is
  analyzed in the small noise approximation, . The atoms are in a two-photon detuning with respect to the
  carriers of the field.  We calculate the stationary quadrature noise
  spectrum of the field outside the cavity in the case where the input probe
  field is a squeezed state and the input pump field is a coherent state. The
  mean value of the field is unaltered in all the analysis: the atoms shows
  electromagnetically induced transparency (EIT). The effect of the atoms'
  base level decoherence in the cavity output field is also studied. It is
  found that the output field is very sensitive to two-photon detuning.
\end{abstract}
\pacs{42.50.Gy,42.50.Lc,42.50.Ar,42.50.Pq}
\maketitle
\vskip2pc

\section{Introduction}
\begin{figure}[t]
  \includegraphics[width=3in]{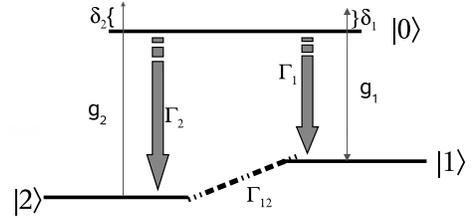}
  \caption{\label{fig:atomo} The atoms are in $\Lambda$ configuration,
    $\Gamma_1$ ($\Gamma_2$) represent the radiative decay constant to state
    $|1\rangle$ ($|2\rangle$). The probe (pump) field has a detuning
    $\delta_2$ ($\delta_1$) with respect to the corresponding dipolar
    transition. $\Gamma_{12}$ accounts for decoherence due to dephasing in the
    atom base levels.}
\end{figure}

It was recently shown that a probe field, after interacting with atoms that
show electromagnetically induced transparency (EIT), does not necessarily preserve
the noise properties of an initially squeezed state. This was first shown for
the stationary output field from a two-mode cavity filled with atoms in
$\Lambda$ configuration (see Fig.~\ref{fig:atomo}) in a fully quantized model.
In addition to the expected reduction of squeezing for frequencies in resonance with
the normal mode splitting of the atom-cavity
system, an interchange of noise properties between the squeezed state probe and
coherent state pump field was found for frequencies for which the mean value
of the field is unaltered (well inside the EIT transparency window and the
cavity bandwidth) \cite{rv:pablo}.  In the case of an initially squeezed state
propagating in an EIT medium, a similar result holds; in this case the
interchange of the noise properties between probe and pump oscillates with
the propagating distance \cite{rv:pablomarc}.  In both cases the transfer of
noise properties is maximum for equal strengths of the probe and pump fields.
These results mean that, although the mean value of the field is unaltered,
the quantum state is indeed altered by the interaction with EIT media. Except
for the expected absorption due to the linewidth of the excited level, the
transfer of noise properties is coherent: the squeezed state is transferred
from the probe to the pump. As there are several proposals for applications
that depend on EIT \cite{rv:lukincollo} \cite{rv:rmpfleisch}
\cite{rv:lasercoolingteo} and because of the coherent nature of the
interaction between the fluctuations of the probe and pump fields in a
quantized model, it is important to understand all the characteristics of the
interaction of quantized fields in EIT.

In this paper we extend the study initiated in \cite{rv:pablo} to include the
case where both cavity modes have the same detuning from the atomic transition
and to analyze the effect of decoherence of the base levels of the atom.
Similarly to the case where both fields are in resonance, when there is
two-photon detuning, the mean value of the field is unaltered. Nevertheless, it is
found that the noise spectra of the output field depend strongly on the
two-photon detuning.  We also studied the effect on the output field of
decoherence in the base levels of the atoms.

This paper is organized as follows. In Sec.~\ref{sc:equation} we give a
brief review of how the system equations are obtained and solved. In Sec. \ref{sc:sqvc} we analyze
the cavity output field for the case where the incoming probe field is a
vacuum squeezed state and the probe mode is not driven. In Sec.~\ref{sc:sq} we
analyze the case where the probe mode is driven. In Sec.~\ref{sc:deco} we study the influence of decoherence of the atom base level on
the cavity output fields. Finally we present our
conclusions in Sec.~\ref{sc:conclusions}.

\section{Cavity output field equations}
\label{sc:equation}

Consider the case of $N$ three-level atoms in $\Lambda$ configuration inside a
cavity that sustains two modes of the electromagnetic field. The annihilation
field operators for each mode are denoted by $\hat a_1$ and $\hat a_2$. Each
mode $\hat a_i$ has a detuning $\delta_i$ with respect to the transitions
between level $|i\rangle$ and $|0\rangle$, $i=1,2$.  We will work with a
two-photon detuning, $\delta_1=\delta_2=\delta$. The atomic system equations are
given by \cite{rv:pablo}

\begin{subequations}
  \label{eq:sistemacav}
  \begin{eqnarray}
    \frac{d}{dt}\hat W_1 &=&\frac{1}{3}(-2 \Gamma_1-\Gamma_2)(1+\hat W_{1}+\hat W_2)-2 i g_1\,\hat\Sigma_{01}\hat a_1 
    \nonumber\\&&+2 i g_1\,\hat a^\dagger_1\hat\Sigma_{10} -i g_2\,\hat\Sigma_{02}\hat a_2 
    \nonumber\\&&+i g_2\,\hat a^\dagger_2\hat\Sigma_{20}  +\hat
    F_{W_1}\, , 
    \\
    \frac{d}{dt}\hat W_2 &=&\frac{1}{3}(-\Gamma_1-2 \Gamma_2)(1+\hat W_{1}+\hat W_2) -i g_1\,\hat \Sigma_{01}\hat a_1 
    \nonumber\\&&+i g_1\,\hat a^\dagger_1\hat \Sigma_{10}  -2 i g_2\,\hat \Sigma_{02}\hat a_2 
    \nonumber\\&&+2 i g_2\,\hat a^\dagger_2\hat \Sigma_{20}  +\hat
    F_{W_2}\, ,
    \\
    \frac{d}{dt}\hat \Sigma_{10}&=&(-\frac{\Gamma_1+\Gamma_2}{2}+i\delta)\hat
    \Sigma_{10}+i g_1\,\hat W_1\hat a_1 \nonumber\\&& -i g_2\,\hat \Sigma_{12}\hat a_2 
    +\hat F_{10}\, ,
    \\
    \frac{d}{dt}\hat \Sigma_{20}&=&(-\frac{\Gamma_1+\Gamma_2}{2}+i\delta)\hat
    \Sigma_{20}+i g_2\,\hat W_2\hat a_2 \nonumber\\&&   -i g_1\,\hat \Sigma_{21}\hat a_1 
    +\hat F_{20}\, ,
    \\
    \frac{d}{dt}\hat{\Sigma}_{21}&=&-\Gamma_{12}\hat{\Sigma}_{21}-i g_1\,\hat a_1^\dagger\hat{\Sigma}_{20}+i
    g_2\,\hat{\Sigma}_{01}\hat a_2 \, ,
  \end{eqnarray}
\end{subequations}
where $\hat{\Sigma}_{ij}=\sum_{k=1}^N\hat{\sigma}_{ij}^k$ are the collective
operators that represent the sum of individual atomic operators
$\sigma_{ij}^k=|i\rangle\langle j|^k$ associated with the $k^{\rm th}$ atom.

Each intracavity mode interacts with its own collection of modes in the
outside field. This means that either they have different polarizations or
their difference in frequency is large.  Using input-output theory
\cite{lb:wallsmilburn} to relate the inside field with the outside field we
obtain the following equations for the intracavity modes \cite{rv:pablo}

\begin{eqnarray}\label{eq:cavidad}
  \frac{d}{dt}\hat a_1(t) & = & -i g_1 \hat \Sigma_{10}(t)-\frac{\gamma_1}{2}
  \hat a_1(t)+\sqrt{\gamma_1}\,\hat a_{1{\rm in}}(t)\, ,\nonumber\\
  \frac{d}{dt}\hat a_2(t) & = & -i g_2 \hat \Sigma_{20}(t)-\frac{\gamma_2}{2}
  \hat a_2(t)+\sqrt{\gamma_2}\,\hat a_{2{\rm in}}(t)\, ,
\end{eqnarray} 
where $\gamma_i$ is the decay rate of cavity mode $i=1,2$.  The operators
$\hat a_{i{\rm in}}(t)=-1/\sqrt{2\pi} \int_{-\infty}^{\infty}d\omega
\, e^{-i\omega(t-t_0)}\hat b_i(t_0,\omega)$ represent the field entering the
cavity. The operator $\hat b_i(t_0,\omega)$ represents the outside mode
associated with cavity mode $i$ at the initial time $t_0$ and frequency
$\omega$.  We will call the outside field associated with the modes labeled by
index $i=1$ ($i=2$) the pump (probe) field. The outcoming field is given by
$\hat a_{i{\rm out}}(t)=1/\sqrt{2\pi} \int_{-\infty}^{\infty}d\omega
\, e^{-i\omega(t-t_1)}\hat b_i(t_1,\omega)$. The operators $\hat b_i(t_1,\omega)$
represent the outside mode associated with the cavity mode $i$ at time $t_1>t_0$
and frequency $\omega$.  All the operators are in a
reference frame rotating with the corresponding cavity frequencies.  The
laboratory frame notation can be obtained by the transformation
$\Sigma_{0j}'={\Sigma}_{0j}\exp{(-i \delta_{j}-\omega_j)}$. Due to the
rotating frame, the spectrum frequency $\omega$ will represent the detuning from
the cavity frequency.  The incoming and outcoming fields are related by

\begin{eqnarray}\label{eq:io}
  \hat a_{1{\rm in}}(t)+\hat a_{1{\rm out}}(t)&=&\sqrt{\gamma_1}\hat a_1(t)\, ,\nonumber\\
  \hat a_{2{\rm in}}(t)+\hat a_{2{\rm out}}(t)&=&\sqrt{\gamma_2}\hat a_2(t)\, .
\end{eqnarray} 

The Langevin fluctuation operators $\hat{F}$'s are assumed to be delta
correlated, with zero mean:
\begin{equation}
  \langle \hat F_x\rangle=0\label{langevin}\, ,
\end{equation}
\begin{equation}
  \langle \hat F_x(t)\hat F_y(t')\rangle=D_{xy}\delta(t-t')\, ,
  \label{eq:langevin}
\end{equation}
where $x$ and $y$ label the fluctuation operators.

The atom diffusion coefficients $D_{xy}$ can be obtained using the
generalized Einstein relations \cite{lb:cohen}. The nonzero diffusion
coefficients are given in \cite{rv:pablo}.

We will consider the following initial conditions for the incoming field. For
frequencies different from the intracavity frequencies, each mode outside the
cavity is, for the probe field, a $\theta=0$ quadrature vacuum squeezed state
and, for the pump field, a vacuum state.  When the frequency is equal to the
probe intracavity frequency, the initial condition is, for the probe field, a
$\theta=0$ squeezed state (mean value different from zero) and for the pump
field a coherent state.

Defining the field quadrature $\theta$ for the field $i$ and frequency
$\omega$ as
\begin{equation}
  \hat Y_{i\, \theta}(\omega,t)=\hat b_i(\omega,t)\exp{(i\theta)}+\hat
  b^\dagger_i(\omega,t)\exp{(-i\theta)}\, ,
\end{equation}
we have that, for the given initial conditions, the $\theta=0$ quadrature
noise operator for the probe field is,
\begin{eqnarray*}
  \Delta Y_{2\,\theta}(\omega,t=t_0)&=&\langle(\hat Y_{i\,\theta}(\omega,t_0)-\langle\hat
  Y_{i\,\theta}(\omega,t_0)\rangle)^2\rangle\\ &=& e^{-2 r}\cos^2\theta+e^{2 r}\sin^2\theta\, ,
\end{eqnarray*}
and $\Delta Y_{1\,\theta}(\omega,t=t_0)=1$ for the pump field, where $r>0$
measures the maximum level of squeezing of the $\theta=0$ quadrature.

That means that the outside modes are initially in a vacuum, which can be
squeezed, except for the resonant modes with the cavity which can be in a
squeezed state but with field mean value different from zero. We will chose
that value in such a way that inside the cavity we have $\langle \hat a_i
\rangle=\alpha_i$.

The system of Eqs.~\eqref{eq:sistemacav}, \eqref{eq:cavidad} and \eqref{eq:io}
is transformed to c-number equations and solved for the stationary case in
the small noise approximation. The method is described in detail in
\cite{rv:davidovich} for two-level systems. A similar system (the resonance
case without decoherence) is discussed in \cite{rv:pablo}. We will suppose
$\Gamma_1=\Gamma_2=\Gamma$ and $\gamma_1=\gamma_2=\gamma$. We will concentrate
on the case $\delta>0$ and $\omega>0$, the negative case being symmetric.

\section{Probe field is a squeezed vacuum}
\label{sc:sqvc}
In this section the probe mode of the cavity is not driven, its mean
value being $\langle\hat a_2\rangle=0$. The pump mode is driven and has $\langle\hat
a_1\rangle=\alpha$.
\subsection{Numerical results}
The results for the output field, although analytical, are to big to give
here. In order to gain insight into the behavior of the cavity output
field we will plot it for different detunings.

\begin{figure}[t]
  \includegraphics[width=3in]{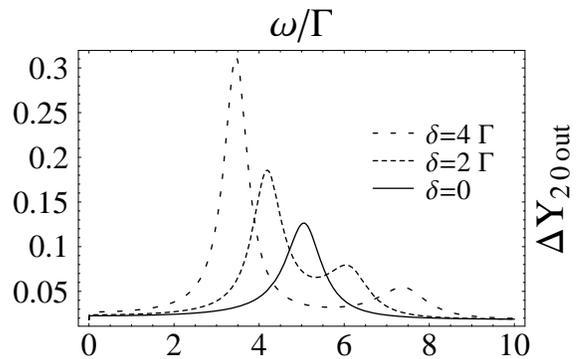}
  \caption{\label{fig:sqvc1} Output probe field fluctuations for the
    $\theta=0$ quadrature in cavity EIT as a function of two photon detuning,
    $\delta_1=\delta_2=\delta$ . The incoming probe field is in a broadband
    squeezed vacuum state. The pump field is in a coherent state. The position
    of one maximum increases with $\delta$ and is due to the expected normal
    splitting of energy in the atom+cavity system. The position of the other
    maximum decreases with increasing $\delta$ and the noise increases. Parameters:
    $g=-0.005\Gamma$, $\Omega_1=\Gamma$, $\Omega_2=0$, $\gamma=.06\Gamma$,
    $N=1000000$, $r=2$.}
\end{figure}

\begin{figure}[t]
  \subfigure{\includegraphics[width=3in]{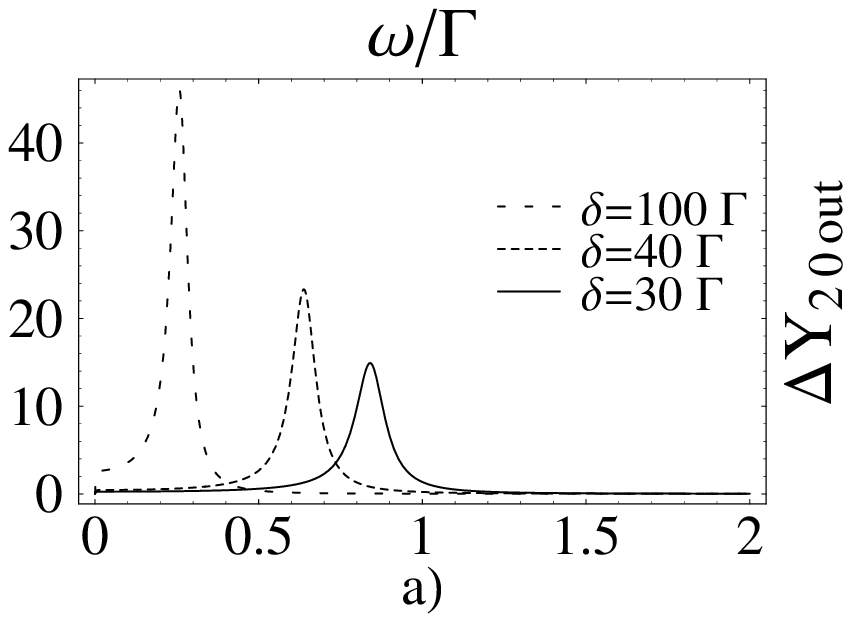}\label{fig:sqvc2a}}\\
  \subfigure{\includegraphics[width=3in]{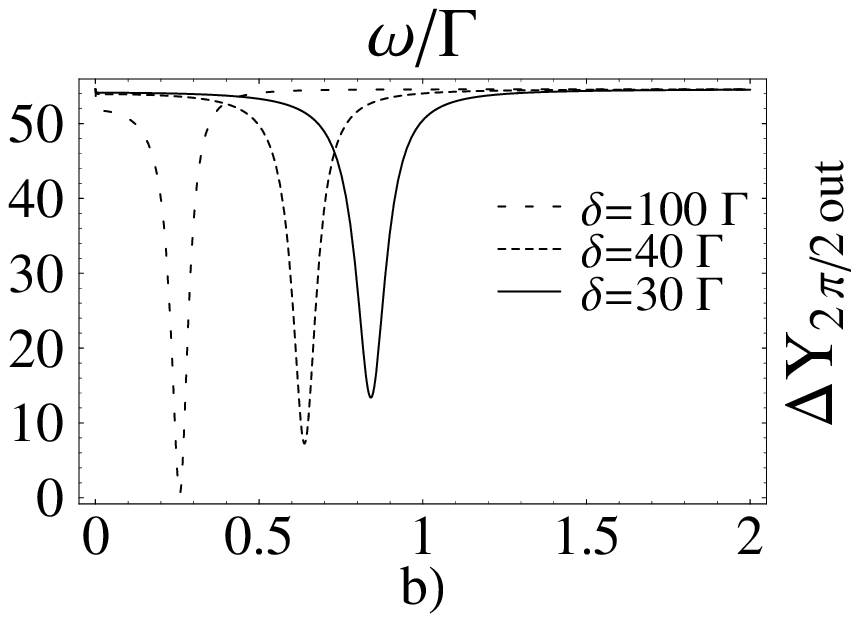}\label{fig:sqvc2b}}
  \caption{\label{fig:sqvc2} Output probe field fluctuations for the
    $\theta=0$ (a) and $\theta=\pi/2$ (b) quadratures in cavity EIT as a
    function of two-photon detuning $\delta$. For $\delta\approx 100\Gamma$
    the minimum in the $\theta=\pi/2$ case is squeezed and corresponds to the
    initial probe condition with the quadrature rotated by $\pi/2$.
    Parameters: $g=-0.005\Gamma$, $\Omega_1=\Gamma$, $\Omega_2=0$,
    $\gamma=.06\Gamma$, $N=1000000$, $r=2$. }
\end{figure}

In Fig. \ref{fig:sqvc1} we plotted the $\theta=0$ quadrature of the field $2$
for different detunings. When $\delta=0$ we obtain a reduction of the incoming
squeezing for some observation frequencies $\omega$. We will call the peak of
these frequencies $\omega_{\delta=0}$. This reduction of squeezing is due to
the normal mode splitting of the atom-cavity system.  When $\delta$ increases,
we observe that the one peak shown for the case $\delta=0$ develops into a
double peak (see the plot for $\delta=2 \Gamma$ and $\delta=4 \Gamma$). One of
the peaks is for an observation frequency $\omega<\omega_{\delta=0}$ and with
a height larger than the other peak for $\omega>\omega_{\delta=0}$. For
$\delta$ even greater, the position of the peak for $\omega<\omega_{\delta=0}$
approaches $\omega=0$ and the maximum of the peaks no longer shows noise
squeezing. This can be seen in Fig.~\ref{fig:sqvc2a}. For the case
$\delta=100\Gamma$ the excess noise is approximately the same as the initial
condition for the $\theta=\pi/2$ quadrature. In Fig.~\ref{fig:sqvc2b} we
plotted the $\theta=\pi/2$ quadrature.  In this plot we can see a reduction of
noise for the same frequencies where we have an increase of noise for the
$\theta=0$ quadrature.  For the case where $\delta=100\Gamma$ the
$\theta=\pi/2$ quadrature of the output field shows squeezing, the numerical
value is approximately the same as the initial condition for the $\theta=0$
quadrature. This result, together with the excess noise shown by the output
field for the $\theta=0$ quadrature, may lead us to suspect that for that frequency we
have a $\pi/2$ rotation of the initial condition.  For frequencies
$\delta<100\Gamma$ we have a similar behavior but the rotation is not
complete.

The maximum position for the peak for $\omega<\omega_{\delta=0}$ increases
with the number of atoms, $N$, and diminishes with increasing two-photon detuning
$\delta$.

\begin{figure}[t]
  \subfigure{\includegraphics[width=3in]{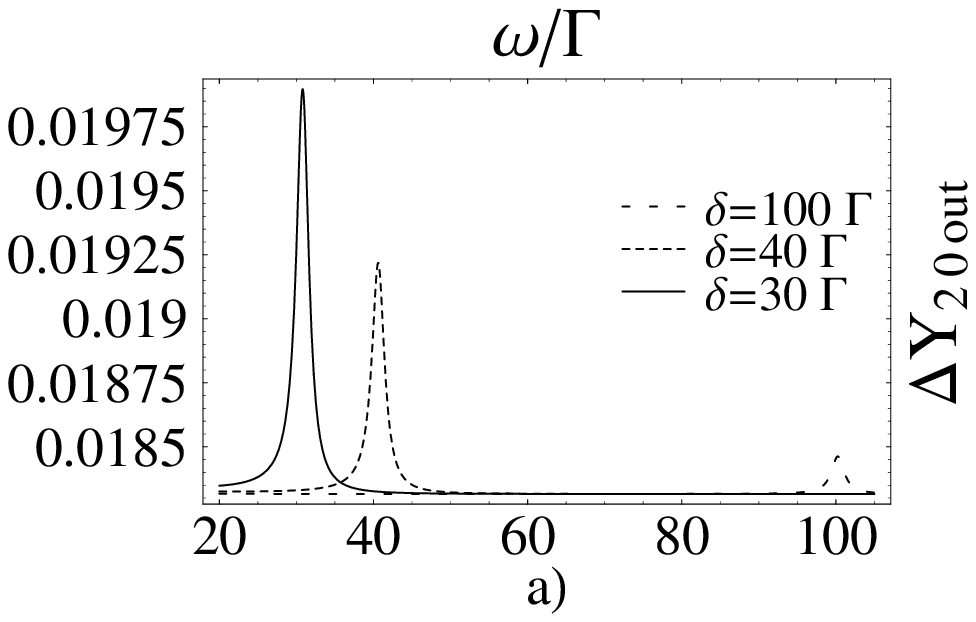}\label{fig:sqvc3a}}\\
  \subfigure{\includegraphics[width=3in]{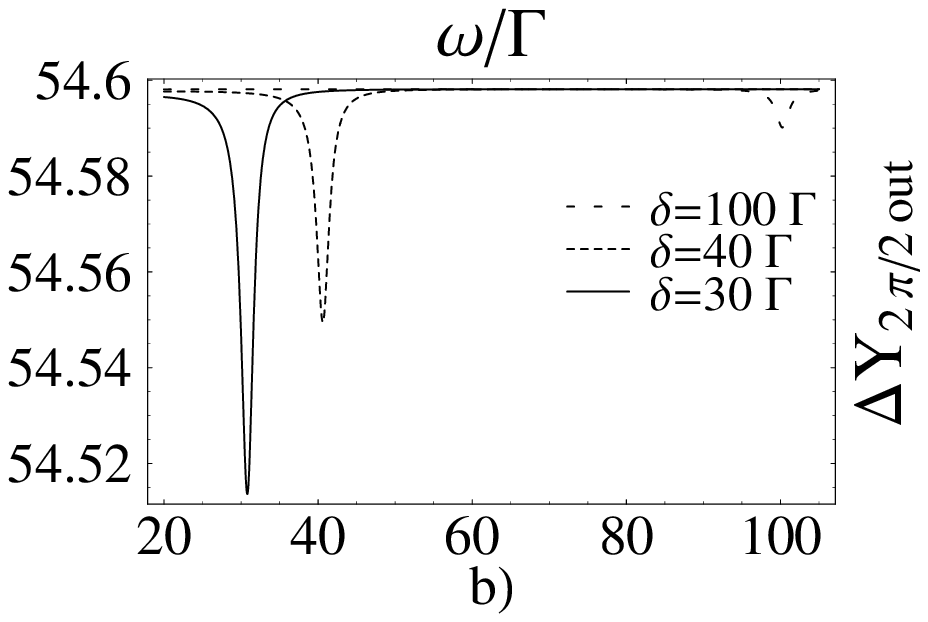}\label{fig:sqvc3b}}
  \caption{\label{fig:sqvc3} Same as Fig.~\ref{fig:sqvc2} for spectrum
    frequency $\omega>20$. It can be seen how for $\delta>30\Gamma$ the effect
    of the normal splitting resonance frequency is very small. Parameters:
    $g=-0.005\Gamma$, $\Omega_1=\Gamma$, $\Omega_2=0$, $\gamma=.06\Gamma$,
    $N=1000000$, $r=2$. }
\end{figure}

In Fig. \ref{fig:sqvc3} we show what happens with the peaks for
$\omega>\omega_{\delta=0}$. We can see that, as $\delta$ increases, the
observation frequency for these peaks also increases and its height
diminishes. The frequency observation of these peaks also increases with $N$.

\subsection{Analytical results}

In order to find analytical expressions for the extremal position, we Taylor
expand the field in $\gamma$. Comparing the term of order $O(\gamma)$ with the
term of order $O(\gamma^2)$, we conclude that the expansion is valid for
$\gamma\ll\Gamma$ and $\omega\geq\sqrt{g^2 N}$. This expansion allows us to
find the position of the extremum for $\omega>\omega_{\delta=0}$. For that
extremum, the peak is located for
\begin{equation}\label{eq:vcsqexp}
  \omega_{>max}=\frac{1}{2} \left(\delta +\sqrt{4 \Omega^2+4 N g^2+\delta
      ^2}\right)\, .
\end{equation}
In the case of $N$ two-level atoms inside a
  cavity, it is known that the absorption spectrum has a peak for frequency $\frac{1}{2} \left(\delta +\sqrt{4 N g^2+\delta
      ^2}\right)$ \cite{rv:agarwal2niveles}. This frequency corresponds to the
    vacuum Rabi splitting for $N$ atoms inside the cavity. As we
    expect that for large detuning of the probe field the three-level
    atoms behave similarly to two-level atoms, we can interpret the frequency
    $\omega_{>max}$ as the usual absorption peak for two-level atoms inside a
    cavity, modified by the presence of an external field (the pump
  field).

In order to obtain analytical results for the position and value of the peak
for $\omega<\omega_{\delta=0}$, we use the numerically observed fact that the
position of this peak increase with $N$ and decrease with increasing $\delta$.  We
write then $\delta=\lambda N$ and take the limit $N\rightarrow\infty$. We
obtain
\begin{eqnarray*}
  \lefteqn{\Delta Y_{2\,\theta_2}(\omega)=\frac{1}{M} \Bigg\{ }\\ & & e^{-2 r} \left(\cos (\theta ) \left(\delta _c^2 \left(4 \omega _{\gamma }^2+1\right)-4\right)-4 \sin (\theta ) \delta
    _c\right)^2\\ & &+e^{2 r} \left(4 \cos (\theta ) \delta _c+\sin (\theta ) \left(\delta _c^2 \left(4 \omega _{\gamma
        }^2+1\right)-4\right)\right)^2\Bigg\}\, ,
\end{eqnarray*}
where
\begin{equation}
  M=16+\left(4 \omega _{\gamma }^2+1\right)^2 \delta _c^4+8\left(1-4
    \omega_{\gamma }^2\right) \delta _c^2\, ,
\end{equation}
$\omega=\omega_\gamma \gamma$, $\delta_c=\delta\, C$ and the cooperativity
parameter $C=g^2 N/\gamma$.  When $\theta=0$ the maximum is located
at
\begin{equation}\label{eq:vcsqmax}
  \omega_{<max}/\gamma=\pm\frac{\sqrt{4-\delta _c^2}}{2 \delta _c}
\end{equation}
if $\delta_c\leq 2$ and $\omega_{<max}=0$ if $\delta_c>2$.  To see the
validity of the former limit for a given $\delta$, we study the expansion of
the output field in the parameter $1/N$. Comparing the order $0$ with the
order $1$ for $\omega_{<max}$, it can be seen that the limit is valid when
$\gamma\ll\Gamma$ and $\delta^2\gg 4 C \Gamma$.

The value of the $\theta$ quadrature for this maximum is

\[
\Delta Y_{2\,\theta_2}(\omega_{<max})=\cos^2(\theta)e^{2
  r}+\sin^2(\theta)e^{-2 r}\, .
\]
The previous equation means that in the limit $N\rightarrow\infty$ the effect
of two-photon detuning in a cavity-atom system is to rotate by $\pi/2$ the
initial condition for the frequency $\omega_{<max}$.

We interpret the peak at frequency $\omega_{<max}$ in the following way. For each
  interaction of the field with the atoms, a small rotation of the maximum
  squeezed quadrature
   happens. This rotation is explained by the phase gained by the field
   induced by the two-photon detuning \cite{rv:fleischstwo-photon}. The
   velocity of this rotation depends on the two-photon detuning and spectrum frequency. The frequency
  positions $\omega_{<max}$ correspond to the frequencies where the buildup of the rotation is maximal. 

In summary, the spectrum at $\omega_{>max}$
  is explained by the energy splitting due to the cavity-atom system (this
  splitting increases with atom number and detuning). The spectrum at
  $\omega_{<max}$ is explained by the phase gained by the field due to
  two-photon detuning.

\section{Probe field is a squeezed state}
\label{sc:sq}
In this section, the incoming squeezed and coherent states which are resonant with the
respective probe and pump modes of the cavity, are such that $\langle\hat
a_2\rangle=\langle\hat a_1\rangle=\alpha$.
\subsection{Numerical Results}
\begin{figure}[t]
  \includegraphics[width=3in]{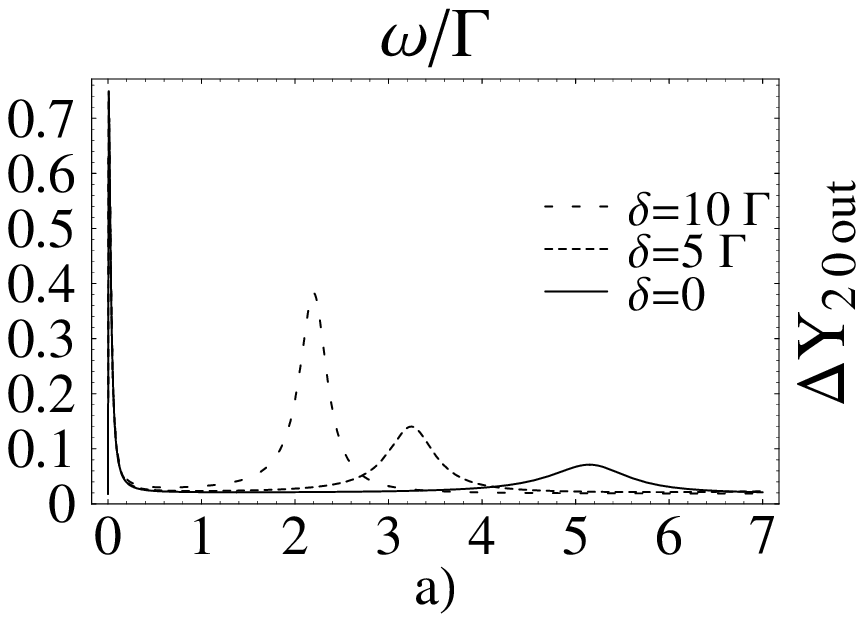}\\
  \includegraphics[width=3in]{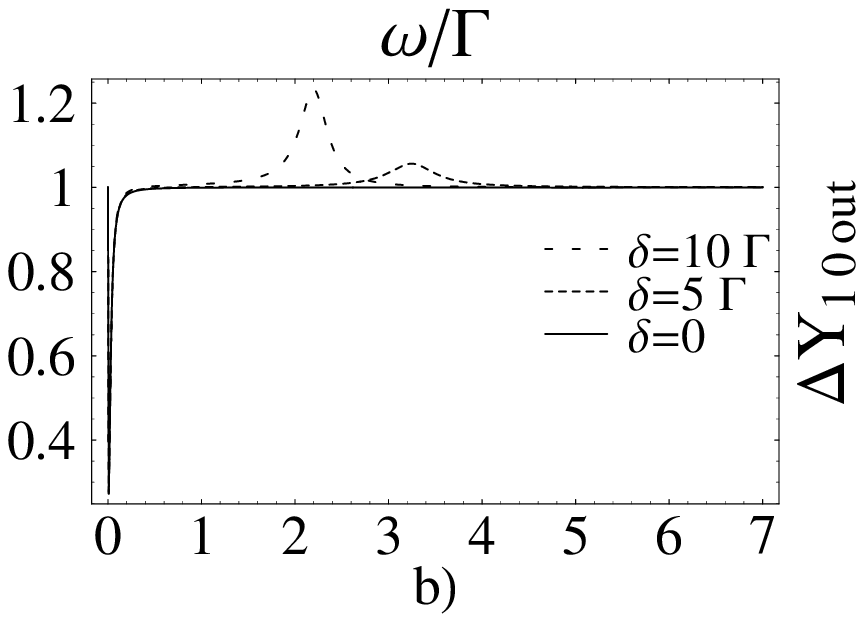}
  \caption{\label{fig:sq1} Output probe (a) and pump (b) field fluctuations
    for the $\theta=0$ quadrature in cavity EIT as a function of two-photon
    detuning $\delta$. The probe field is in a broadband vacuum squeezed state
    except
    for the resonance mode with the cavity. The pump field is in a coherent
    state. For $\omega$ in the vicinity of zero, it can be observed how the
    transfer of squeezing between the probe and pump fields does not depend on
    $\delta$. For $\omega/\Gamma>1$, a maximum, whose frequency decreases with
    increasing $\delta$, can be seen in the pump and probe output fields.
    Parameters: $g=-0.005\Gamma$, $\Omega_1=\Omega_2=\Gamma$,
    $\gamma=.06\Gamma$, $N=1000000$, $r=2$.}
\end{figure}

\begin{figure}[t]
  \includegraphics[width=3in]{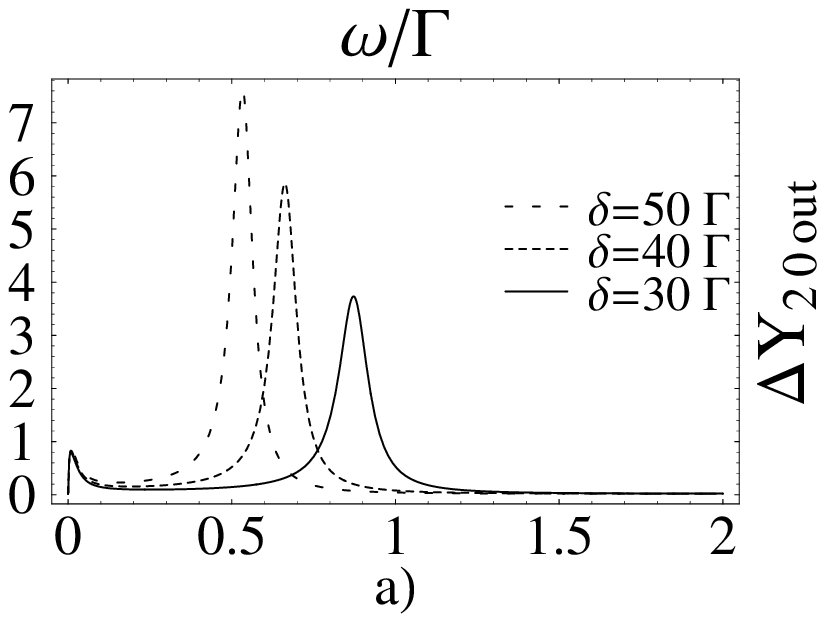}\\
  \includegraphics[width=3in]{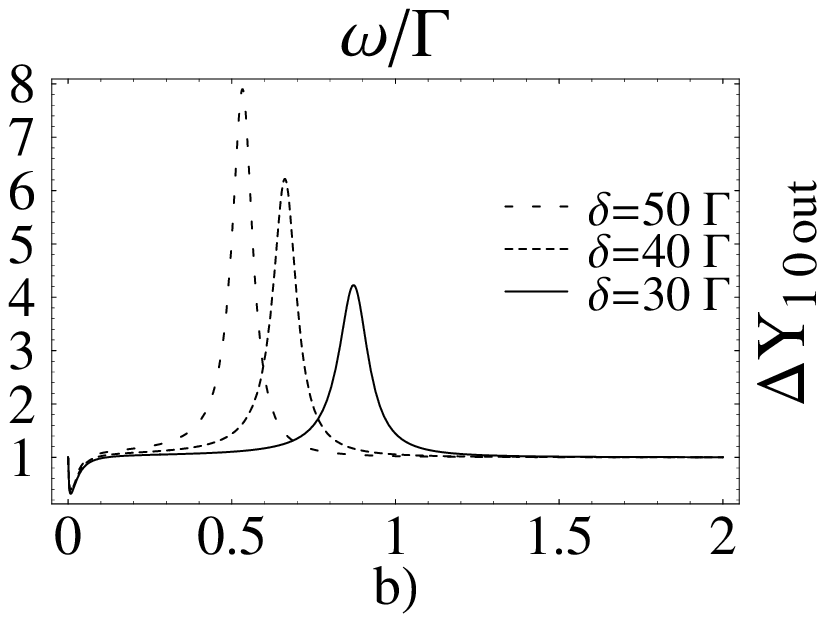}
  \caption{\label{fig:sq2} Same as Fig.~\ref{fig:sq1} for
    $\delta=30\Gamma,40\Gamma,50\Gamma$.}
\end{figure}

In Figs. \ref{fig:sq1} and \ref{fig:sq2} we plot the $\theta=0$ and
$\theta=\pi/2$ quadratures of the output field. When $\omega/\Gamma>1/4$ the
extrema found have a qualitatively similar behavior to that studied in the
previous section, namely, the extremum found for $\delta=0$ around
$\omega/\Gamma\approx 5$ divides into two as $\delta$ increases. The location of
one of these extrema increase with $\delta$ (not shown in the plot) and the
other decreases. We will focus on the case of spectrum frequency
$\omega<\omega_{\delta=0}$. Unlike in the squeezed vacuum case, we can
see from the figures that the pump field also has an increase of noise. For
$\delta\gg 1$ the maximum noise is given by approximately $e^{4}/4$ instead of
$e^{4}$ as in the vacuum squeezed case. In the next section we will obtain
 an analytical expression, valid for some parameters, for the positions and values
of these extrema.  For $\omega/\Gamma<1/4$ we observe the interchange of
squeezing described in \cite{rv:pablo} for $\delta=0$. It can be seen that,
for the parameters in the figures, this interchange of squeezing apparently
does not depend on the two-photon detuning $\delta$. We found that this is
true until $\delta$ is so large that the excess noise peak enters the domain
of the frequencies where we had the interchange of squeezing. In the next
section we will better characterize this behavior.

\subsection{Analytical Results}

In order to find analytical expressions for the extremal positions, we Taylor
expand the field in $\gamma$. Comparing the term of order $O(\gamma)$ with the
term of order $O(\gamma^2)$, we conclude that the expansion is valid for
$\gamma\ll\Gamma$ and $\omega\geq\sqrt{g^2 N}$. This expansion allows us to
find the position of the extremum for $\omega>\omega_{\delta=0}$. For that
extremum, the peaks are located at
\[
\omega_{>max}=\frac{1}{2} \left(\delta +\sqrt{8 \Omega^2+4 N g^2+\delta
    ^2}\right)\, .
\]

In order to obtain analytical results for the positions and values of the peaks
for $\omega<\omega_{\delta=0}$, we use the numerically observed fact that the
positions of these peaks increase with $N$ and decrease with increasing $\delta$.  We
write then $\delta=\lambda N$ and take the limit $N\rightarrow\infty$. For the
$\theta=0$ quadrature we obtain (the expression for all quadratures is too
large to write here)
\begin{eqnarray*}
  \lefteqn{\Delta Y_{2\,\theta_2=0}(\omega)=\frac{1}{R} \Bigg\{ }\\ & & 4 e^{2 r} \left(4 \omega _{\gamma }^2+1\right) \delta _c^2+ 4 \left(\left(4 \omega _{\gamma }^2+1\right) \delta
    _c^2+4\right)+\\ & &e^{-2 r} \left(\left(4 \omega _{\gamma }^2+1\right)^3 \delta _c^4-32 \left(4 \omega _{\gamma }^4+\omega
      _{\gamma }^2\right) \delta _c^2+64 \omega _{\gamma }^2\right)\Bigg\}\, ,
\end{eqnarray*}
where
\[
R=\left(1+4 \omega_\gamma ^2\right) M\, .
\]

To see the validity of the former limit for a given $\delta$, we study the
expansion of the output field in the parameter $1/N$. Comparing the order $0$
with the order $1$ for $\omega_{<max}$, it can be seen that the limit is valid
when $\gamma\ll\Gamma$, $\delta^2\gg 4 C \Gamma$, and $C\gamma\gg \Omega^2$.

When the term proportional to $e^{2 r}$ is much larger that the other two, the
spectral frequency for the extremum is given by Eq.~\eqref{eq:vcsqmax}.
Substituting this extremum in the expression for the quadratures of the field,
$\Delta Y_{i\,\theta}(\omega_{<max})$, we obtain

\begin{eqnarray}\label{eq:sqmax}
  \lefteqn{\Delta Y_{2\,\theta}(\omega_{<max})=\Delta
    Y_{1\,\theta+\pi/2}(\omega_{<max})=}\nonumber\\ & &\frac{1}{4}
  \left(\left(e^{-r}+e^r\right)^2+\left(-e^{-2 r}+e^{2 r}\right) \cos (\theta )
    \sin (\theta ) \delta _c\right)\, .\nonumber\\
\end{eqnarray}

The previous equation, valid for $\delta_c\leq 2$, summarizes the most interesting
behavior of the interaction of a broadband probe squeezed state with a pump
coherent state in cavity EIT. The characteristics of the stationary output
field, for spectral frequency $\omega_{<max}$, are as follows. (i) The probe maximum
squeezing quadrature is rotated, from the $\theta=0$ quadrature, corresponding
to the initial condition, to $\theta=3\pi/4$. (ii) The probe field is no
longer a minimal uncertainty state. (iii) The pump field is also squeezed. In
fact, the quadrature spectrum of the pump field is equal to the quadrature
spectrum of the probe field rotated by $\pi/2$. We explain these results
as a combination of quadrature noise rotation due to two-photon detuning, as
explained in Sec.~\ref{sc:sqvc}, and interchange of noise properties
between pump and probe, as explained in \cite{rv:pablo}.

It is known \cite{rv:pablo} that the frequency where the exchange of squeezing
happens is $\omega_{sq}=\gamma g\alpha/(\sqrt{2}\sqrt{g^2 N
  /\Gamma+2g^2\alpha^2})$. We can guess then that, as long as $\delta$ is such
that $\omega_{max}\gg\omega_{sq}$, the interchange of squeezing for
frequencies inside the cavity linewidth is unaffected by the two-photon
detuning. We could not find an analytical probe of this hypothesis but we
checked it numerically for the parameters in the figures.

We explain now the transfer of squeezing, between probe and pump fields,
  that takes place in the spectral
  frequency $\omega_{sq}$. In order to understand this phenomenon, we recall
  what happens in the case of Gaussian states propagating in
  EIT media \cite{rv:pablomarc}. For this case, there is a transfer of
  squeezing between the pump and probe field as a function of spectral
  frequency and propagation distance. When the spectral frequency is the same
  as that of the driven fields, nothing happens. This the reason why, in the cavity,
  although the field interacts several times with the atoms before leaving the
  cavity, nothing happens for $\omega=0$. When the spectral frequency is detuned
  with respect to the driven frequency, transfer of squeezing starts to
  happen. What we obtain for the cavity output field is the transfer of
  squeezing that builds up during the several times the field interacts with
  the medium as it bounces between the cavity mirrors. As the spectral frequency
  increases, the transfer of squeezing increases, but at the same time, due to
  the frequency width of the cavity ($\gamma$), the mode associated with this frequency interacts less with
  the atoms. The frequency where the maximum of the transfer of squeezing
  happens, $\omega_{sq}$, is the result of an equilibrium between both
  processes. For small detuning, the distance scale of the rotation of the
  maximum squeezed quadrature is larger than the scale of the transfer of
  squeezing. This explains why small detuning does not affect the buildup of
  the transfer of squeezing. 

As noted before, the effect of the quantum field-atom interaction, for the
output field spectrum at frequency $\omega_{<max}$, given by Eq.~\eqref{eq:sqmax}, can be viewed as the combination
of rotation of the quadrature of maximum squeezing with interchange of
noise properties between probe and pump fields.
\section{Decoherence of atom base levels}
\label{sc:deco}
\subsection{Probe is a squeezed vacuum}

When the probe is a squeezed vacuum, the mean value of the output field is the
same as the mean value of the input field and the mean values of the dipole
operators are zero. So there is no fluorescence. The decoherence between the
base levels of the atoms ($ \Gamma_{12}\neq 0$) affects only the statistical
properties of the incoming field, namely, degrading the maximum squeezed
quadrature. In order to see this we start discussing the case where the
spectral frequency is the same as the two-photon detuning ($\omega=0$).
Substituting $\alpha_2=0$ but allowing $\Gamma_{12}\neq 0$ in the system
equations, we found the following expression for the $\theta=0$ quadrature:
\begin{eqnarray}
  \lefteqn{\Delta Y_{2\,\theta_2=0}(\omega=0)=\frac{1}{B^2} \Bigg( 16 e^{2
      r} C^2 \delta ^2 \Gamma _{12}^4+}\nonumber\\ & & e^{-2 r} \left(\Omega^4-4 C^2 \Gamma _{12}^2+2 \Omega^2  \Gamma  \Gamma_{12}+\right.\nonumber\\ & & \left. \left(\Gamma
      ^2+\delta ^2\right) \Gamma _{12}^2\right)^2\Bigg )+ \frac{1}{B} 8 C
  \Gamma _{12} \left(\Omega^2+\Gamma  \Gamma _{12}\right)\label{eq:vcpr}\, ,
\end{eqnarray}
where
\[
B=\delta ^2 \Gamma _{12}^2+\left( \Omega^2+2 C \Gamma _{12}+\Gamma \Gamma
  _{12}\right)^2\, .
\]
The $\pi/2$ quadrature, $\Delta Y_{2\,\theta_2=\pi/2}$, has a similar
expression but with $r$ substituted by $-r$.

Due to the term $e^{2 r}$, we identify the first term of Eq.~\eqref{eq:vcpr}
with a $\pi/2$ rotation of part of the noise of the $\pi/2$ quadrature initial
condition. It is known that the field emitted from a medium has a phase of
$\pi/2$ with respect to the incoming field. The fact that part of the $\pi/2$
quadrature incoming noise is rotated to $\theta=0$ noise is already known
from the fluorescence of two-level atoms \cite{rv:walserzoller}.  The novelty
here is that there is no fluorescence (the mean value of the field is zero)
but there is still transformation of noise from the $\pi/2$ quadrature to the
$\theta=0$ quadrature.  This transformation of noise is proportional to the
two-photon detuning, being zero when $\delta=0$.  The other two terms of
Eq.~\eqref{eq:vcpr} are different from zero when $\delta=0$; we
interpret them as degradation of squeezing due to the interaction of the
squeezed state in media subject to decoherence.

For $\omega>\gamma$ we could not find numerically any significant difference
from the plots of Sec.~\ref{sc:sqvc} due to decoherence in the base levels.

\subsection{Probe is a squeezed state}

Here we analyze the influence of the atom base level decoherence
($\Gamma_{12}\neq 0$) in the cavity output field. We compare our result
with the cavity output field studied in Sec.~\ref{sc:sq}.

We could not find analytical solutions; we carry out our study numerically. The
parameters not specified in this section are the same used in Sec.~\ref{sc:sq}.

For $\omega>\gamma$ we could not find any significant difference from the
plots in Sec.~\ref{sc:sq} due to decoherence in the base levels. The
biggest values of $\Gamma_{12}$ we use are approximately the ones that turn
the system equations unstable. For the parameters in the plots, this value is
around $\Gamma_{12}=\Gamma/210$. This means that the special characteristics
found in the previous section for spectrum frequency $\omega_{<max}$, namely
quadrature rotation and noise interchange between probe and pump, are not
sensitive to base level decoherence, as long as $\omega_{<max}>\gamma$.

\begin{figure}[t]
  \subfigure{\includegraphics[width=3in]{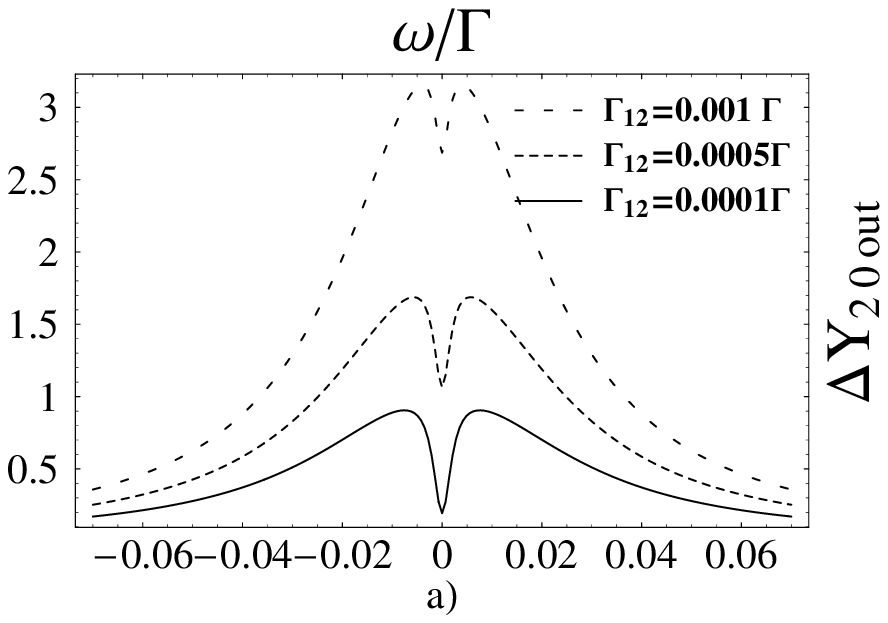}\label{fig:gammap}}\\
  \subfigure{\includegraphics[width=3in]{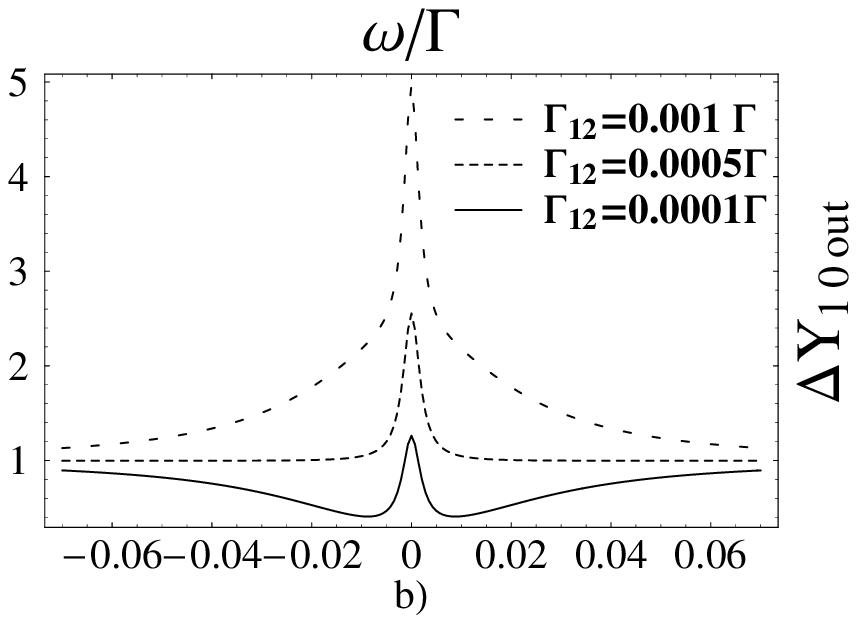}\label{fig:gammab}}
  \caption{\label{fig:gamma} Output probe (a) and pump (b) field
    fluctuations for the $\theta=0$ quadrature in cavity EIT as a function of
    base level decoherence $\Gamma_{12}$. The probe field is in a broadband
    vacuum squeezed state except for the resonance mode with the cavity. The pump
    field is in a coherent state. It can be seen how the transfer of squeezing
    between the probe and pump field is very sensitive to
    $\Gamma_{12}$.Parameters: $g=-0.005\Gamma$, $\Omega_1=\Omega_2=\Gamma$,
    $\gamma=.06\Gamma$, $N=1000000$, $r=2$. }
\end{figure}

For $\omega<\gamma$ we found that the transfer of squeezing between the probe
and pump is destroyed for very small $\Gamma_{12}$. This is shown in
Fig.~\ref{fig:gamma} (compare with Figs.~\ref{fig:sq1} and \ref{fig:sq2} for
$\omega<\gamma$). As can be seen in the figure, for $\Gamma_{12}$ of the order
of $\Gamma_{12}=0.0005\Gamma$, the pump quadrature [Fig. \ref{fig:gammab}] does
not show any squeezing.  We explain this in the following way. As
$\alpha_2=\alpha_1$, the effect of decoherence is to break the dark state and
some fluorescence comes from the atom (the emitted field is different from
zero \cite{rv:pablonicim}). The cavity forces the atoms to emit photons inside
its linewidth. This fluorescence destroys the noise properties of the squeezed
state for frequencies inside the cavity linewidth ($\omega<\gamma$).

\begin{figure}[t]
  \subfigure{\includegraphics[width=3in]{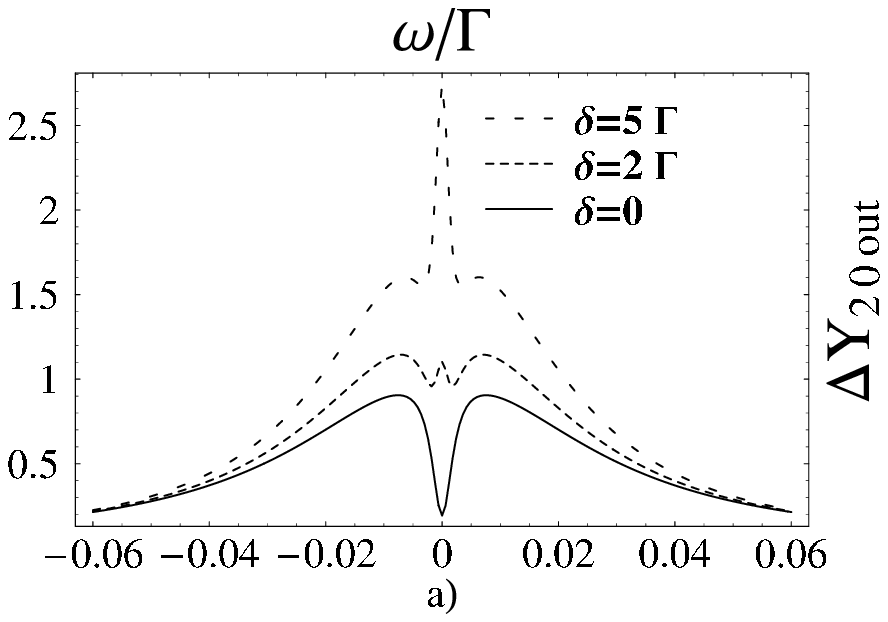}\label{fig:deltap}}\\
  \subfigure{\includegraphics[width=3in]{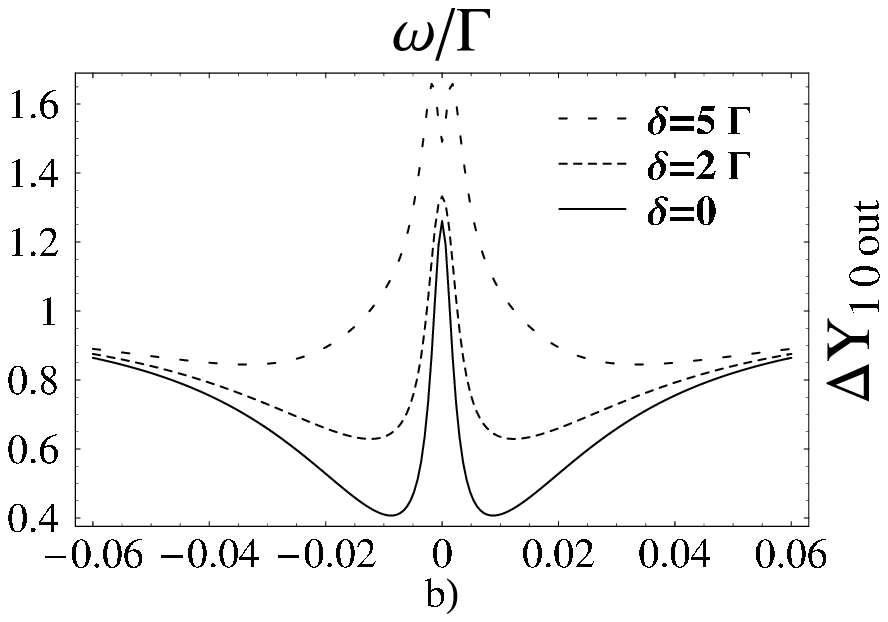}\label{fig:deltab}}
  \caption{\label{fig:delta}Output probe (a) and pump (b) field fluctuations
    for the $\theta=0$ quadrature in cavity EIT as a function of two-photon
    detuning $\delta$ when there is decoherence in the atom base levels.
    Unlike the case $\Gamma_{12}=0$, here it can be seen how the output
    field strongly depends on two photon detuning. Parameters:
    $g=-0.005\Gamma$, $\Omega_1=\Omega_2=\Gamma$, $\gamma=.06\Gamma$,
    $N=1000000$.$\Gamma_{12}=\Gamma/10000$ }
\end{figure}

The two-photon detuning also has a huge effect in destroying the transfer of
squeezing between the probe and pump fields. We show this in
Fig.~\ref{fig:delta}.  We use a value of the decoherence rate
($\Gamma_{12}=\Gamma/10000$) for which the transfer of squeezing is almost
unaffected for $\delta=0$. We plot then the probe and pump $\theta=0$ noise
spectra for $\delta=0,2\Gamma,5 \Gamma$. Remember that for $\Gamma_{12}=0$ the
cases $\delta=2\Gamma, 5\Gamma$ would be similar to the $\delta=0$ case. It can be observed in the
figure how, unlike in the case $\Gamma_{12}=0$, the noise strongly depends on
$\delta$, and increases for all frequencies if $\delta$ increases.

\section{Summary and Conclusions}
\label{sc:conclusions}
We have studied the output field of a two-mode cavity sustaining two-photon
detuning ($\delta_1=\delta_2=\delta$; see Fig. \ref{fig:atomo}) with $N$ atoms
in $\Lambda$ configuration. The atoms show EIT, and the mean value of the field
is practically unaltered by the interaction of the cavity+atom system.
Nevertheless, there is a coherent alteration of the incoming noise properties
in the output field. This alteration depends strongly on the two-photon
detuning and the mean value of the modes inside the cavity. Decoherence tends
to destroy the quantum properties of the state (i.e., the squeezing) for
frequencies inside the cavity linewidth. The particular characteristics of the
output field are as follows.

In the case of an incoming coherent pump and a vacuum squeezed probe, the
output pump field remains coherent but the output probe field suffers large
alterations with respect to the incoming broadband vacuum squeezed state.
In addition to a frequency [see~Eq.\eqref{eq:vcsqexp}] where absorption of
squeezing is expected due to normal splitting of the cavity+atom system, we
found that, for a spectral frequency which decreases with increasing $\delta$
[see Eq.\eqref{eq:vcsqmax}], the quadratures of the output probe field shows a
partial rotation of the incoming quadrature probe field. In some cases,
$\delta^2\gg 4 C \Gamma$, this rotation can be complete and the incoming
$\theta=0$ quadrature squeezed field is transformed into a $\theta=\pi/2$
quadrature squeezed field. The effect of decoherence is, in addition to
degrading the
incoming vacuum squeezed state, to rotate part of the incoming $\theta=\pi/2$
quadrature noise to the $\theta=0$ quadrature for frequencies inside the
cavity linewidth.

In the case where the cavity probe mode is also pumped and both modes inside
the cavity have the same strength, both modes suffer large alteration with
respect to the incoming quantum state. In the case of $\omega>\gamma$, the
output field has a maximum which decreases with increasing $\delta$.
Nevertheless we no longer have a rotation of the quadrature of the
incoming field as in the previous case.  When $\delta^2\gg 4 C \Gamma$ and
$C\gamma\gg\Omega^2$, the characteristics of the stationary output field, for
spectral frequency $\omega_{<max}$, are that (i) the probe maximum squeezing
quadrature is rotated, from the $\theta=0$ quadrature, corresponding to the
initial condition, to $\theta=3\pi/4$; (ii) the probe field is not longer a
minimal uncertainty state; and (iii) the pump field is also squeezed. In fact, the
quadrature spectrum of the pump field is equal to the quadrature spectrum of
the probe field rotated by $\pi/2$. We explain these results as a
combination of quadrature noise rotation due to two-photon detuning and
interchange of noise properties between pump and probe.

When $\omega<\gamma$ the transfer of fluctuation between probe and pump is
unaltered by the two-photon detuning as long as $\omega_{max}>\gamma$.
Nevertheless, the combination of decoherence and two-photon detuning has a
large effect in the field fluctuation for frequencies inside the cavity
linewidth, destroying the squeezed state for small $\Gamma_{12}$.

Our results shows, that interaction of quantum fields in cavity EIT is a much richer phenomeno than believed, and that the quantum nature of both fields, probe and pump, should be taken into account when dealing with cavity EIT.

\acknowledgments

We thank CONACYT for financial support. Part of this work was done during a
workshop in the Centro Internacional de las Ciencias (CIC).


\end{document}